\documentclass[preprint]{aastex}

\begin{document}

\title {The ${\it Kepler}$ Eclipsing System KIC 5621294 and its Substellar Companion}
\author{Jae Woo Lee$^{1,2}$, Kyeongsoo Hong$^1$, and Tobias Cornelius Hinse$^1$}
\affil{$^1$Korea Astronomy and Space Science Institute, Daejeon 305-348, Korea}
\affil{$^2$Astronomy and Space Science Major, Korea University of Science and Technology, Daejeon 305-350, Korea}
\email{jwlee@kasi.re.kr, kshong@kasi.re.kr, tchinse@gmail.com}

\begin{abstract}
We present the physical properties of KIC 5621294 showing light and timing variations from the ${\it Kepler}$ photometry.
Its light curve displays partial eclipses and O'Connell effect with Max II fainter than Max I, which was fitted quite well
by applying third-body and spot effects to the system. The results indicate that the eclipsing pair is 
a classical Algol-type system with parameters of $q$=0.22, $i$=76$^\circ$.8, and $\Delta$($T_{1}$--$T_{2}$)=4,235 K, in which 
the detached primary component fills about 77\% of its limiting lobe. Striking discrepancies exist between the primary and 
secondary eclipse times obtained with the method of Kwee \& van Woerden. These are mainly caused by surface inhomogeneities 
due to spot activity detected in our light-curve synthesis. The 1,253 light-curve timings from the Wilson-Devinney code were 
used for a period study. It was found that the orbital period of KIC 5621294 has varied due to a periodic variation overlaid 
on a downward parabola. The sinusoidal variation with a period of 961 d and a semi-amplitude of 22.5 s most likely arise from 
a light-time effect due to a third component with a mass of $M_3 \sin i_3$=46.9 M$\rm_{Jup}$, which is in good agreement with
that calculated from the light curve itself. If its orbital inclination is larger than about 40$^\circ$, the mass of 
the circumbinary object would possibly match a brown dwarf. The parabolic variation could not be fully explained by either 
a mass transfer between the binary components or an angular momentum via magnetic braking. It is possible that the parabola 
may only be the observed part of a period modulation caused by the presence of another companion in a wider orbit. 
\end{abstract}

{}

\section{INTRODUCTION}

Theoretical studies have proposed that a significant number of close binaries form as the inner binaries in triple systems 
and that the tertiary components play an important role in the formations of the tight pairs, presumably through energy and 
angular momentum exchanges (Eggleton \& Kisseleva-Eggleton 2006; Fabrycky \& Tremaine 2007). In such hierarchical triples, 
a binary orbit has shrunk from an initially wider dimension by transferring most of the angular momentum of the central binary 
to the distant circumbinary companion through dynamical interaction. Then, the close binary with a short orbital period should 
evolve into a contact configuration by an angular momentum loss (AML) due to magnetic braking (Bradstreet \& Guinan 1994). 
The observational evidence for this scenario was made in the statistical study by Pribulla \& Rucinski (2006), finding that 
59$\pm$8\% of W UMa contact binaries have companions. Tokovinin et al. (2006) showed that a strong correlation exists between 
the binary period and the existence of a third companion and found that nearly all close binaries ($P<3.0$ d) reside in triple 
and higher systems. These results imply that the circumbinary object is necessary for the production of a short-period binary. 

It is possible to precisely measure the conjunction instants of eclipse binaries from their light curves. Physical phenomena 
associated with the timing measurements include mass exchange or loss, angular momentum loss, apsidal motion in an eccentric orbit, 
light-travel-time (LTT) effect, and magnetic cycle of active stars (Hilditch 2001). The LTT effect caused by a third body occurs 
because the relative distance changes as the eclipsing binary moves around the barycenter of the multiple system (Irwin 1952). 
Thus, the presence of the tertiary component causes a cyclical variation of the eclipse times of a binary star. We can 
potentially discover multiple systems with circumbinary companions by searching for periodic features in the observed ($O$) $-$ 
calculated ($C$) minimum epochs {\it versus} time. The detection of such systems is of great interest because the masses and 
orbits of the companions can be directly determined from the $O$--$C$ diagrams alone. If long historical timing data have 
precisions on the order of a few seconds, the eclipsing binaries showing an LTT effect offer us an opportunity for detecting 
substellar companions, such as giant planets and brown dwarfs (Lee et al. 2009; Hinse et al. 2012; Pribulla et al. 2012). 
However, the orbital periods for many well-studied binaries have varied by a combination of some causes, rather than in 
a monotonic fashion (for examples, see Kreiner et al. 2001\footnote {http://www.as.up.krakow.pl/o-c/}). Moreover, 
timing variations with small amplitudes could be produced by the asymmetries of eclipse light curves due to stellar activity 
such as starspots and pulsations (Kalimeris et al. 2002; Gies et al. 2012; Lee et al. 2014c). Eclipsing binaries provide 
significant information on the dynamics of the systems and their structure and evolution through the timing data. 

The {\it Kepler} satellite provides the exquisitely precise and uninterrupted observations over a long time baseline for 
many eclipsing binaries. We have been looking for circumbinary objects orbiting binary stars from whole analyses of 
the light curves and eclipse timings, based on all available data including the {\it Kepler} observations. Our early results 
have led to the discoveries of a triply eclipsing hierarchical triple (KIC 2856960; Lee et al. 2013) and a quadruple system 
exhibiting a substellar companion and $\gamma$ Dor pulsations (V404 Lyr; Lee et al. 2014b). In order to advance this subject 
further, we choose the {\it Kepler} target KIC 5621294 (2MASS J19285262+4053359), which was announced to be a detached eclipsing
variable with an orbital period of 0.9389 d by Pr\v sa et al. (2011). In this paper, we present and discuss the discovery of 
a substellar object in the KIC 5621294 system from both the light-curve synthesis and the eclipse timing analysis.

\section{LIGHT-CURVE SYNTHESIS AND ABSOLUTE DIMENSIONS}

The {\it Kepler} photometry of KIC 5621294 has been performed during Quarters 1$-$11 and 14$-$17 in long cadence mode, 
corresponding to a sampling rate of 29.4 min. From the total dataset, 57,510 individual observations were used for our study 
after removing eight points, which showed an unreasonably large deviation compared to neighboring ones. As shown in Fig. 1, 
the {\it Kepler} light curve is superficially similar to that of Algol and displays the conventional O'Connell effect with 
Max II fainter than Max I by about 0.005 mag. The effect is usually interpreted as spot activity on the stellar photospheres 
of the components. In order to determine the geometrical structure and the physical parameters of KIC 5621294, 
the {\it Kepler} data were analyzed in a manner identical to that for the eclipsing system V404 Lyr (KIC 3228863) exhibiting 
LTT effects and multiperiodic pulsations (Lee et al. 2014b). We set the mean light level at phase 0.25 to unity and used 
the 2007 version of the Wilson-Devinney synthesis code (Wilson \& Devinney 1971, van Hamme \& Wilson 2007; hereafter W-D). 
The light-curve syntheses were repeated until the correction of each parameter become smaller than its standard deviation 
using the differential correction (DC) program of the W-D code. The errors for the fitted parameters were estimated following 
the procedure described by Koo et al. (2014).

For modelling the light curve of KIC 5621294, the effective temperature ($T_1$) of the primary star was assumed to be 8,425 K 
from the {\it Kepler} Input Catalogue (Brown et al. 2011). The logarithmic bolometric ($X_{1,2}$) and monochromatic ($x_{1,2}$) 
limb-darkening coefficients were interpolated from the values of van Hamme (1993). The gravity-darkening exponents were fixed 
at standard values of $g_1$=1.0 and $g_2$=0.32, while the bolometric albedos were $A_1$=1.0 and $A_2$=0.5, as surmised from 
the components' temperatures. A synchronous rotation for both components ($F_{1,2}$=1.0) and the detailed reflection effect 
were considered. Accordingly, the adjustable parameters were the mass ratio ($q$), the orbital inclination ($i$), the temperature 
($T_2$) of the secondary star, the dimensionless surface potential ($\Omega_{1,2}$), and the monochromatic luminosity ($L_{1}$). 
Because the {\it Kepler} observations have a time span of about four yr, we can study the orbital period change of the system from 
its light curve independently of the minimum timings discussed in the following section. Thus, we included the orbital ephemeris 
($T_0$, $P$, and $dP$/$dt$) and third-body parameters ($a^{\prime}$, $i^{\prime}$, $e^{\prime}$, $\omega^{\prime}$, $P^{\prime}$, 
and $T_{\rm c}^{\prime}$) as free parameters. Furthermore, spot models were added to fit the asymmetrical light curve and 
the third light source ($\ell_3$) was considered throughout our analyses. But, the trials for a possible $\ell_3$ failed to 
achieve convergence. In this paper, we refer to the primary and secondary stars as those being eclipsed at Min I (at phase 0.0) 
and Min II, respectively, and the primed notation denotes the third-body orbit in the W-D synthesis code. 
The parameter $a^{\prime}$=$a_{\rm b}+a_{3}$, and $T_{\rm c}^{\prime}$ is not the time of periastron passage but the time of 
the superior conjunction in the binary's motion about the mass center of the triple system.

No radial velocity measures have been made for KIC 5621294. Therefore, we calculated a series of models with varying $q$, 
permitting no perturbations such as a third body or a spot. In this $q$-search procedure, 1,000 normal points were formed 
from the {\it Kepler} data and weights were assigned to the number of individual observations per normal point. The results 
for various modes of the W-D code showed acceptable photometric solutions only for mode 5. The behavior of the weighted sum 
of the squared residuals ($\Sigma W(O-C)^2$; hereafter $\Sigma$) as a function of $q$ is displayed in Fig. 2, showing 
a minimum value around $q$=0.22. This indicates that the eclipsing pair is in a semi-detached configuration and permits 
some mass transfer from the less massive secondary to the detached primary component. In subsequent calculations, we used 
all individual points, and treated the mass ratio and third-body parameters as adjustable variables. The unspotted solution 
is given in columns (2)--(3) of Table 1 and the residuals from it are displayed in the middle panel of Fig. 1. 
As shown in the figure, this model does not describe the observations near phase 0.75 because we have established 
the unit light level at phase 0.25. 

In semi-detached Algols, such a feature can be reasonably modelled by including a hot spot on the photosphere of the primary as 
a result of the impact due to a gas stream coming from the secondary star (Lee et al. 2010). However, this idea does not exclude 
the possible existence of a magnetic cool spot located on the surface of the low-mass secondary with a spectral type later than 
about F5 (Hall 1989; Lee et al. 2014b). In the case of KIC 5621294, because the cool secondary with a convective atmosphere 
may be a magnetically active star, we tested two possible spot models: a hot spot on the primary component and a cool spot on 
the secondary. Final results are given in columns (4)--(7) of Table 1. The synthetic curve from the hot-spot model is displayed 
as the solid curves in the top panel of Fig. 1 and the residuals from it are plotted in the bottom panel. From the table and 
figure, we can see that the hot-spot model greatly improves the light-curve fitting and give smaller values of $\Sigma$ than do 
the unspotted model. However, it is difficult to distinguish between the hot- and cool-spot models because there are 
no $\Sigma$ differences between them. In all the procedures that have been described, we considered an orbital eccentricity ($e$) 
for the eclipsing binary but found that the parameter retained a value of zero.

Absolute dimensions for the binary components can be estimated from the photometric solutions and the empirical relations between 
spectral type and stellar mass. The effective temperature of the primary component corresponds to a normal main-sequence star with 
a spectral type of about A4. We computed the absolute parameters for the eclipsing pair listed in the bottom of Table 1. 
The luminosity ($L$) and bolometric magnitudes ($M_{\rm bol}$) were obtained by adopting $T_{\rm eff}$$_\odot$=5,780 K and 
$M_{\rm bol}$$_\odot$=+4.73 for solar values. For the absolute visual magnitudes ($M_{V}$), we used the bolometric corrections (BCs) 
from the relation between $\log T_{\rm eff}$ and BC (Torres 2010). From its spectral type, the primary star would be a candidate 
for $\delta$ Sct or $\gamma$ Dor pulsators. Gies et al. (2012) suggested that the {\it Kepler} light curve of KIC 5621294 displays 
rapid variability related to the pulsation of the primary star. Accordingly, we applied multiple frequency analyses to 
the light curve residuals from our spot models, but found no periodicity with a semi-amplitude larger than 1 mmag.

\section{ECLIPSE TIMING VARIATION AND ITS IMPLICATIONS}

We determined 1,804 times of minimum light and their errors from the {\it Kepler} data using the method of Kwee \& van Woerden 
(1956, hereafter KW). These are given in Table 2, where we present the epochs and $O$--$C$ values computed with 
the linear ephemeris for the unspotted model in Table 1. The $O$--$C$ diagram for the timings is shown in Fig. 3, where 
the filled and open circles represent the primary and secondary minima from the KW method, respectively. In the figure, 
the $O$--$C$ residuals from both types of eclipse are not consistent with each other. Because the eclipsing binary of KIC 5621294
is in a circular orbit, apsidal motion can be ruled out. Then, this discrepancy can be produced by asymmetrical eclipse minima 
due to stellar activity such as starspots and/or the method of measuring the timings of minima (Lee et al. 2009, 2014c). 
The light-curve synthesis method developed by W-D is capable of extracting the times of conjunction and gives better information 
with respect to the KW method, which does not consider spot activity and is based on observations during minimum alone. 
As the light curves of KIC 5621294 were satisfactorily modeled, we calculated minimum times with the W-D code by only adjusting 
both the reference epoch ($T_0$) and spot parameters in each spot model of Table 1. We chose the weighted means of the two values 
from each light curve as our light-curve timings. The results are listed in Table 2 and are illustrated with the plus symbols 
in Fig. 3. As seen in the figure, the W-D timings of 1,253 conformed to the general trend of the primary minima obtained with 
the KW method, but are not matched with the secondary ones at all. 

Inspecting Fig. 3 in detail, we found that the orbital period of KIC 5621294 has varied in a sinusoidal way superposed on 
a downward parabolic variation, rather than in a monotonic fashion such as a parabola or a sinusoid. The sinusoidal term was 
provisionally identified as an LTT effect suggested by the light-curve synthesis. Thus, the W-D timings were finally fitted to 
a quadratic {\it plus} LTT ephemeris, as follows: 
\begin{eqnarray}
C = T_0 + PE + AE^2 + \tau_{3},
\end{eqnarray}
where $\tau_{3}$ is the LTT due to a third body in the system (Irwin 1952, 1959) and includes five parameters ($a_{\rm b}\sin i_3$, 
$e_{\rm b}$, $\omega_{\rm b}$, $n_{\rm b}$, and $T_{\rm b}$). Here, $a_{\rm b} \sin i_3$, $e_{\rm b}$ and $\omega_{\rm b}$ are 
the orbital parameters of the close binary around the mass center of the triple system. The parameters $n_{\rm b}$ and $T_{\rm b}$ 
denote the Keplerian mean motion of the mass center of the binary components and its epoch of periastron passage, respectively.
The Levenberg-Marquart (LM) technique (Press et al. 1992) was applied to solve for the eight unknown parameters of the ephemeris 
and the results are listed in Table 4 together with other related quantities. In this table, $P_{\rm b}$ and $K_{\rm b}$ indicate 
the cycle length and semi-amplitude of the LTT orbit, respectively, and $f(M_{3})$ is the mass function of the system. 

The LM method produces formal errors computed from the best-fitting covariance matrix and they can sometimes be unrealistic 
due to the strong correlations between the model parameters. In order to determine the parameter errors, we performed tens of 
thousands of Monte Carlo bootstrap-resampling experiments (Press et al. 1992, Lee et al. 2014a). For each bootstrap dataset, 
we used the best-fitting parameters as our initial values. In Fig. 4, we plotted the histograms for four main parameters 
($a_{\rm b}\sin i_3$, $e_{\rm b}$, $P_{\rm b}$, and $K_{\rm b}$) as obtained from the 50,000 simulations, where we can see 
that each parameter follows a Gaussian distribution. The parenthesized values in Table 4 are the standard deviations calculated 
from the 50,000 element bootstrap array using the best-fitting model. Within errors, most parameters from the eclipse timings are 
in good agreement with those obtained from the light curve using the W-D code. 

The $O$--$C$ diagram of KIC 5621294 constructed with the linear terms in Table 4 is drawn in the top panel of Fig. 5, where 
the solid curve and the dashed parabola represent the full contribution and the quadratic term, respectively. The middle panel 
refers to the LTT orbit and the bottom panel shows the residuals from the complete ephemeris. These appear as 
$O$--$C_{\rm full}$ in the fourth column of Table 3. As indicated by the figure, the quadratic {\it plus} LTT ephemeris gives 
a satisfactory representation of the ensemble of the timing residuals. The mass function of the third body becomes 
$f_3 (M)$=1.52 $\times10^{-5}$ $M_\odot$ and its projected mass is $M_{3} \sin i_{3}$=0.0448 $M_\odot$, corresponding to 
46.9 M$\rm_{Jup}$ where M$\rm_{Jup}$=M$_\odot$/1047. If the orbital inclination of $i_3$ is larger than about 40$^\circ$, 
the mass of the circumbinary object is below the theoretical threshold of $\sim$0.07 M$_\odot$ for a hydrogen-burning star 
and it would be a brown dwarf.

In eclipse timing diagrams, another possible mechanism for sinusoidal period variations is a magnetic activity cycle for 
systems with a late-type secondary star (Applegate 1992). However, the LTT period (2.633 yr) of KIC 5621294 is too short 
for the magnetic cycle when compared with a median value around 40--50 yr in Algol and RS CVn systems (Lanza \& Rodono 1999).
Moreover, it may be difficult for the Applegate model to produce a strictly periodic variation in the timing residuals. 
These mean that this mechanism cannot contribute significantly to the observed period change of KIC 5621294 and 
that the short-term oscillation most likely arises from the LTT effect due to the existence of a substellar companion 
physically bound to the eclipsing pair. Because the circumbinary object is in a relatively close orbit with the inner binary, 
the timing variations could be affected by the dynamic effect added to the geometrical LTT effect 
(Borkovits et al. 2003; \" Ozdemir et al. 2003). We computed the semi-amplitude of the dynamic perturbation on the motion of 
the eclipsing pair to be less than 0.00001 d and found that its contribution is not significant. 

The negative coefficient of the quadratic term ($A$) listed in Table 4 indicates a continuous period decrease with a rate 
of $-$2.693($\pm$0.006) $\times 10^{-6}$ d yr$^{-1}$, which corresponds to a fractional period change of 
$-$7.372$\times$10$^{-9}$ and, in principle, could be interpreted as a mass transfer from the more massive primary to 
the secondary star or as an angular momentum loss. The secular variation cannot be produced just by mass transfer between 
the components, because the eclipsing pair of KIC 5621294 is a semi-detached binary with a lobe-filling secondary component 
and the secondary to primary mass transfer should increase the orbital period. Accordingly, the parabolic variation cannot 
be interpreted uniquely as mass transfer between the components and its possible cause may be AML due to magnetic stellar 
wind braking in a cool secondary with a convective envelope. The period decrease rate due to spin-orbit-coupled AML has 
been given by Guinan \& Bradstreet (1988), as follows:
\begin{equation}
 {dP \over dt} \approx -1.1 \times 10^{-8} q^{-1} (1+q)^2 (M_{\rm 1}+M_{\rm 2})^{-5/3} k^2 (M_{\rm 1} R^4 _{\rm 1} + M_{\rm 2} R^4 _{\rm 2} ) P^{-7/3},
\end{equation}
where $k^2$ is the gyration constant. With $k^2$=0.1 typical for low-mass main sequence stars (see Webbink 1976), we get 
a theoretical rate of $-$6.52$\times$10$^{-8}$ d yr$^{-1}$ for the magnetic braking, which is over an order of magnitude too small 
to be the single cause of the observed period decrease. We think the parabola may be a part of a longer periodic variation.

\section{SUMMARY AND DISCUSSION}

In this paper, we have presented the photometric properties of KIC 5621294 from the detailed analyses of the light curve and 
the eclipse timings, based on the {\it Kepler} public data. The {\it Kepler} light curve, displaying partial eclipses and 
the O'Connell effect, was satisfactorily modeled by including third-body parameters and by adopting a hot spot on 
the primary component or a cool spot on the secondary. The light-curve synthesis represents that the eclipsing pair of 
the system is an Algol-type semi-detached binary with a mass ratio of $q$=0.22, an orbital inclination of $i$=76$^\circ$.8, 
and a temperature difference between the components of $\Delta$($T_{1}$--$T_{2}$)=4,235 K. The primary component fills 
its limiting lobe by about 77\% and is about 1.4 times larger than the lobe-filling secondary. No pulsating feature was 
detected in the light residuals after subtracting the synthetic eclipsing curves of the spot models from the {\it Kepler} data.

Apparent shifts of the KW secondary minima, shown in Fig. 3, dominantly originate from the spot activity on either of 
the components. Based on the light-curve timings determined from the W-D code, the orbital period of KIC 5621294 has varied 
as a beat effect due to the combination of a downward parabola and a sinusoidal variation, with a period of 961 d and 
semi-amplitude of 22.5 s. The result is in excellent agreement with that calculated from the light curve itself. 
The sinusoidal variation would be produced by the LTT effect due to a gravitationally-bound third body with a minimum mass 
of $M_3 \sin i_3$=0.0448 M$_\odot$ in an eccentric orbit of $e_3$=0.434. The circumbinary companion would be a brown dwarf 
if the orbital inclination is $i_3 \ga$ 40$^\circ$. The decreasing rate of the secular period is calculated to be 
$-$2.693$\times$10$^{-6}$ d yr$^{-1}$, which could be interpreted as AML due to magnetized stellar winds in a cool secondary. 
However, the observed rate is about 42 times larger than the value expected for the magnetic braking and so the AML hypothesis 
is not confirmed. Alternatively, it is possible that the parabolic variation may only be the observed part of another modulation 
with a cycle of about five yr and a semi-amplitude of about 100 s, most likely a second LTT effect caused by the presence of 
a fourth component. 

The results presented in this paper demonstrate that KIC 5621294 is a multiple system, which consists of an eclipsing pair 
with a period of 0.9389 d and at least a substellar companion with an LTT period of about 961 d in an eccentric orbit. 
The existence of the circumbinary components and the dynamics of this system may provide a significant clue to the formation 
and evolution of an initial tidal-locked progenitor of the eclipsing pair by transferring angular momentum via 
Kozai oscillation (Kozai 1962; Pribulla \& Rucinski 2006) or a combination of the Kozai cycle and tidal friction 
(Fabrycky \& Tremaine 2007). This would cause KIC 5621294 to evolve into its present configuration by AML through magnetic braking 
and ultimately into a contact configuration. When high-precision spectroscopic observations are undertaken, all of this 
will be better understood than now and the absolute dimensions and evolutionary status of this system will be advanced greatly. 
A large number of future accurate timings is required to identify the existence of the supposed fourth component.

\acknowledgments{ }
This paper includes data collected by the {\it Kepler} mission. {\it Kepler} was selected as the 10th mission of the Discovery Program. 
Funding for the {\it Kepler} mission is provided by the NASA Science Mission directorate. We have used the Simbad database maintained 
at CDS, Strasbourg, France. This work was supported by the KASI (Korea Astronomy and Space Science Institute) grant 2014-1-400-06.

\clearpage
\begin{figure}
\includegraphics[]{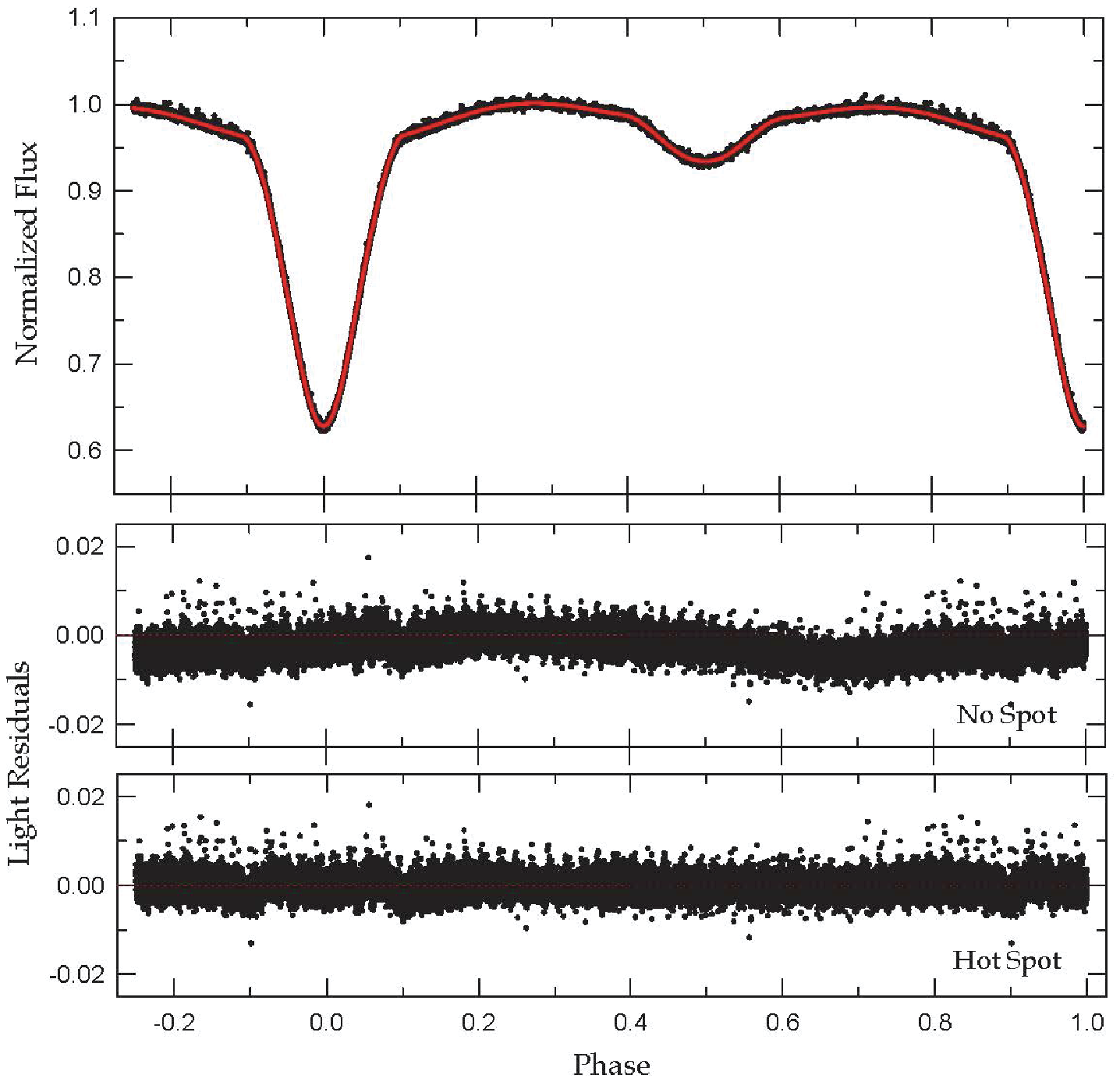}
\caption{Light curve of KIC 5621294 with the fitted model. The circles are individual measures from the {\it Kepler} spacecraft 
and the solid curve is computed with a hot spot on the primary star. The light residuals corresponding to the unspotted and 
hot-spot models are plotted in the middle and bottom panels, respectively. }
\label{Fig1}
\end{figure}

\begin{figure}
\includegraphics[]{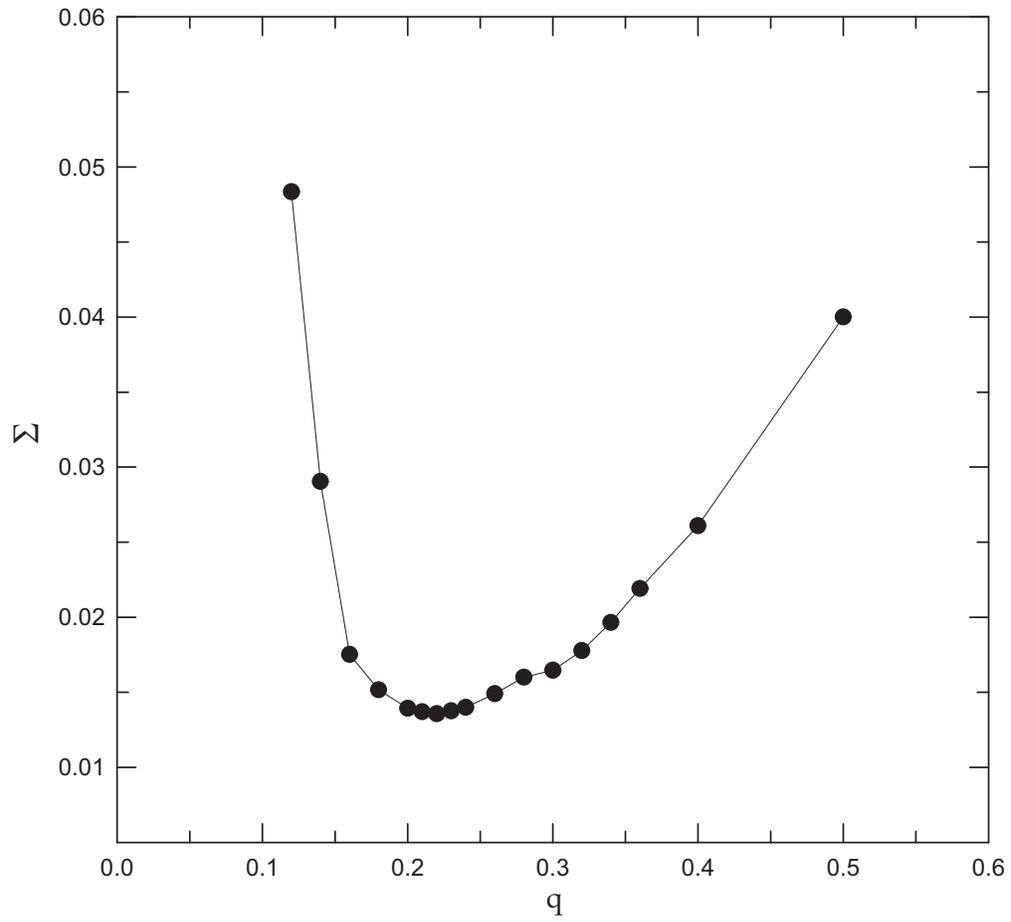}
\caption{Behavior of $\Sigma$ (the sum of the residuals squared) of KIC 5621294 as a function of mass ratio $q$, showing 
a minimum value at $q$=0.22. }
\label{Fig2}
\end{figure}

\begin{figure}
\includegraphics[]{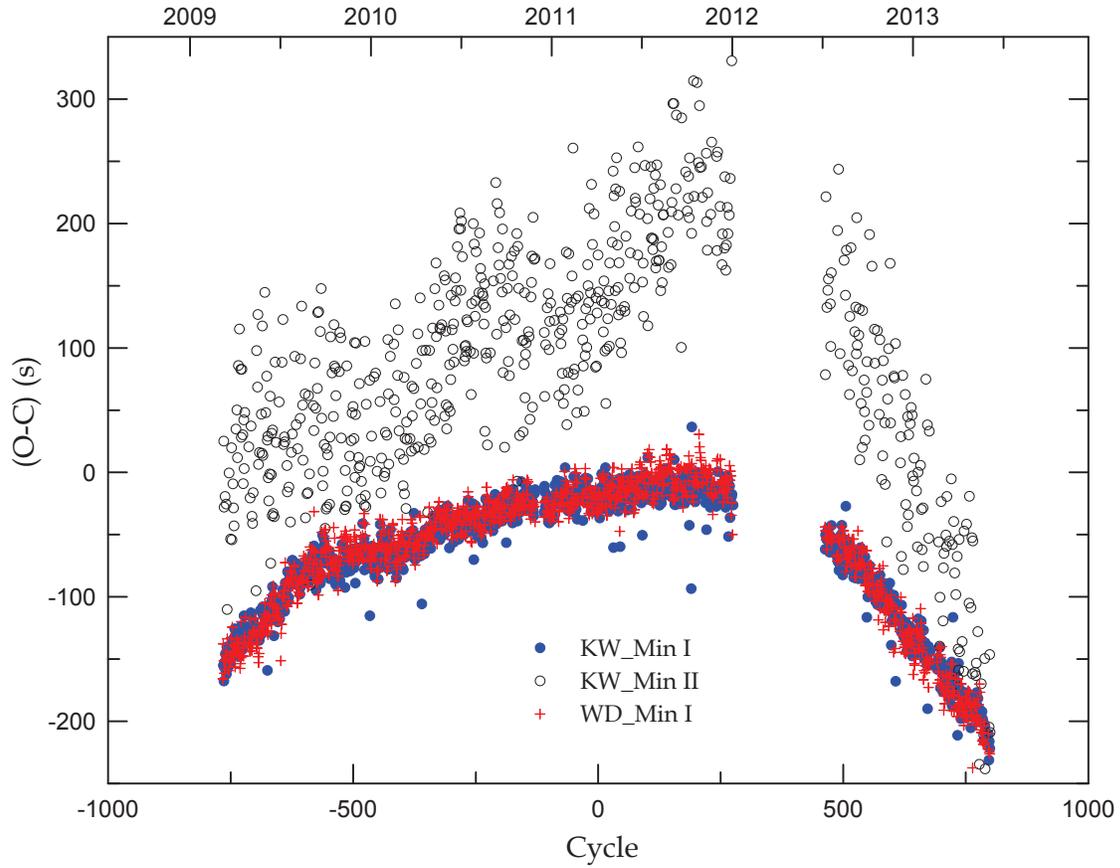}
\caption{$O$--$C$ diagram of KIC 5621294. The filled and open circles represent the primary and secondary minima from the method 
of Kwee \& van Woerden, respectively. The plus symbols refer to the light-curve timings obtained with the W-D code. Because of 
the high density of the points, many of the filled circles cannot be seen individually. }
\label{Fig3}
\end{figure}

\begin{figure}
\includegraphics[scale=0.7]{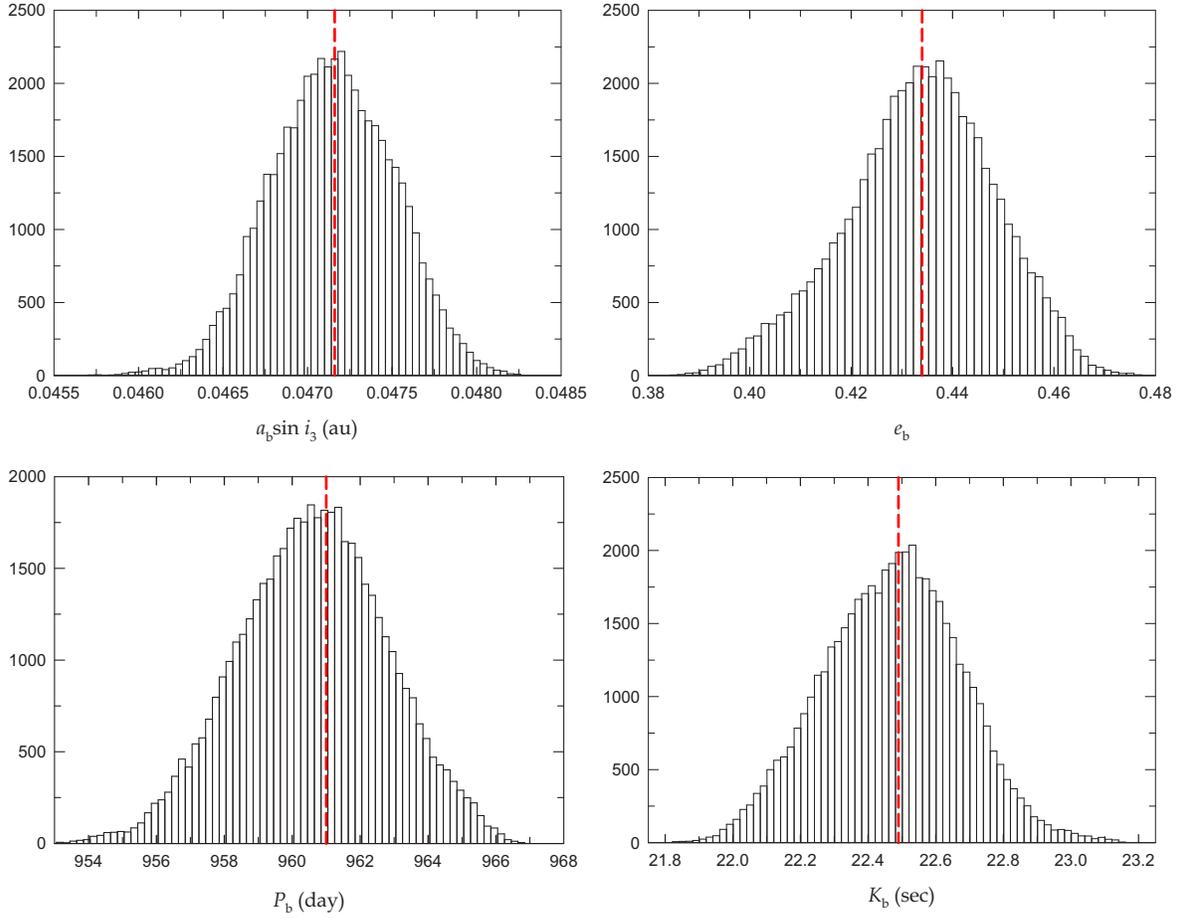}
\caption{Histogram distribution of four main parameters as obtained from 50,000 bootstrap-resampling simulations. In each panel,
the vertical dashed line indicates the location of the best-fitting value derived from the LM method in Table 4. }
\label{Fig4}
\end{figure}

\begin{figure}
\includegraphics[]{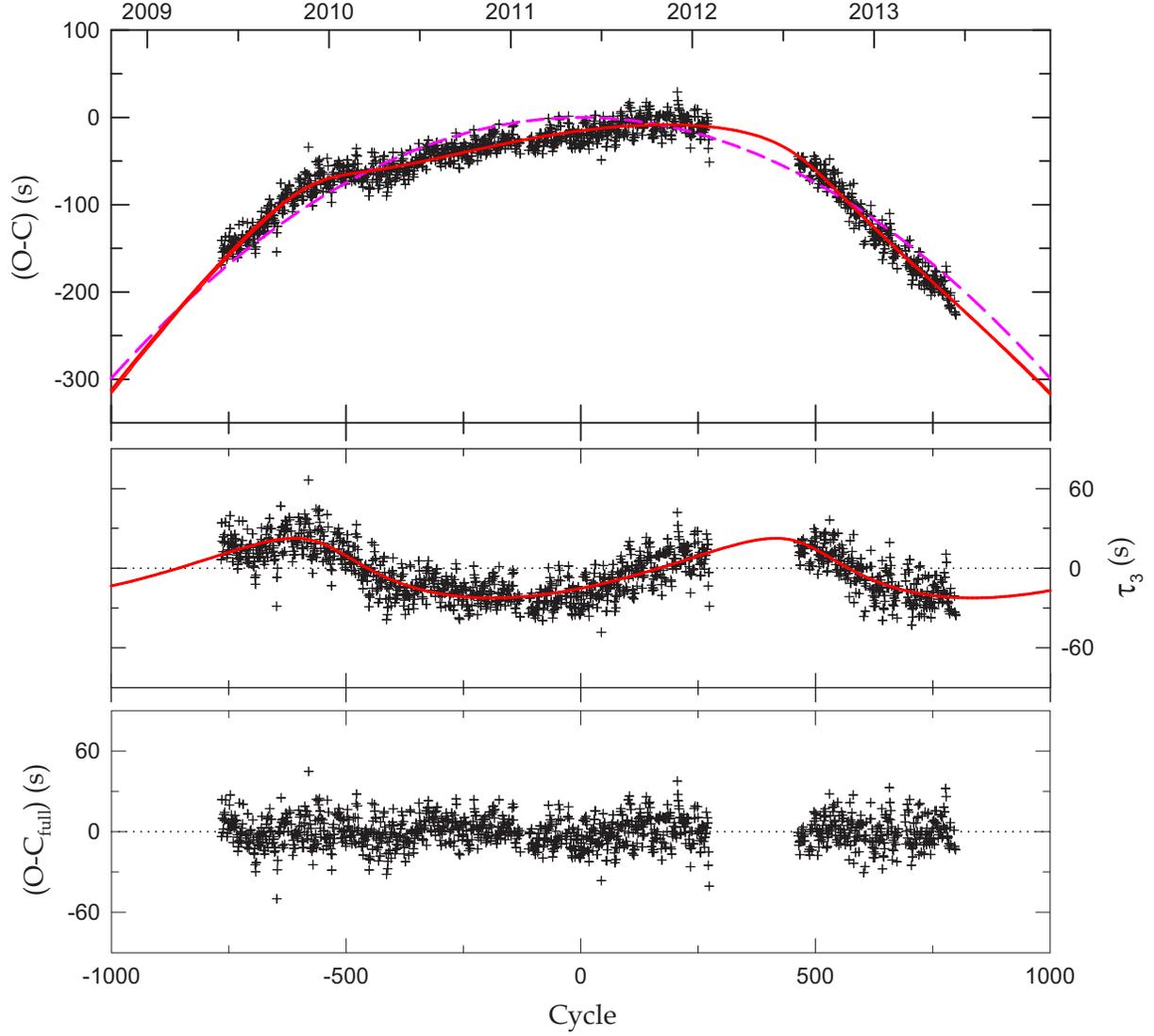}
\caption{In the top panel the $O$--$C$ diagram of KIC 5621294 is constructed with the linear terms in Table 4. The solid and 
dashed curves represent the full contribution and just the quadratic term of equation (1), respectively. The middle panel 
displays the LTT orbit and the bottom panel the residuals from the complete ephemeris. }
\label{Fig5}
\end{figure}

\clearpage
\begin{deluxetable}{lcccccccc}
\tabletypesize{\scriptsize} 
\tablewidth{0pt} 
\tablecaption{Physical parameters of KIC 5621294}
\tablehead{
\colhead{Parameter}                      & \multicolumn{2}{c}{Unspotted Model}             && \multicolumn{2}{c}{Hot-Spot Model}              && \multicolumn{2}{c}{Cool-Spot Model}               \\ [1.0mm] \cline{2-3} \cline{5-6} \cline{8-9} \\[-2.0ex]
                                         & \colhead{Primary} & \colhead{Secondary}         && \colhead{Primary} & \colhead{Secondary}         && \colhead{Primary} & \colhead{Secondary} 
}
\startdata                                                                                                                                                                                         
$T_0$ (BJD)                              & \multicolumn{2}{c}{2,455,672.77636(5)}          && \multicolumn{2}{c}{2,455,672.77648(5)}          && \multicolumn{2}{c}{2,455,672.77655(5)}            \\
$P$ (d)                                  & \multicolumn{2}{c}{0.93890518(2)}               && \multicolumn{2}{c}{0.93890524(2)}               && \multicolumn{2}{c}{0.93890525(2)}                 \\
d$P$/d$t$                                & \multicolumn{2}{c}{$-$7.7(2)$\times$10$^{-9}$}  && \multicolumn{2}{c}{$-$7.8(1)$\times$10$^{-9}$}  && \multicolumn{2}{c}{$-$8.0(1)$\times$10$^{-9}$}    \\
$q$                                      & \multicolumn{2}{c}{0.2215(1)}                   && \multicolumn{2}{c}{0.2144(1)}                   && \multicolumn{2}{c}{0.2256(1)}                     \\
$i$ (deg)                                & \multicolumn{2}{c}{76.796(6)}                   && \multicolumn{2}{c}{76.991(6)}                   && \multicolumn{2}{c}{76.733(6)}                     \\
$T$ (K)                                  & 8,425             & 4,183(5)                    && 8,425             & 4,193(7)                    && 8,425             & 4,186(6)                      \\
$\Omega$                                 & 2.9854(9)         & 2.2854                      && 2.9738(8)         & 2.2683                      && 2.9851(8)         & 2.2953                        \\
$X$, $Y$                                 & 0.661, 0.154      & 0.624, 0.153                && 0.661, 0.154      & 0.624, 0.152                && 0.661, 0.154      & 0.624, 0.153                  \\
$x$, $y$                                 & 0.557, 0.247      & 0.750, 0.074                && 0.557, 0.247      & 0.750, 0.075                && 0.557, 0.247      & 0.750, 0.074                  \\
$l$/($l_{1}$+$l_{2}$)                    & 0.9706(2)         & 0.0294                      && 0.9708(2)         & 0.0292                      && 0.9703(2)         & 0.0297                        \\
$r$ (pole)                               & 0.3601(1)         & 0.2398(1)                   && 0.3607(1)         & 0.2376(1)                   && 0.3606(1)         & 0.2411(1)                     \\
$r$ (point)                              & 0.3812(1)         & 0.3507(1)                   && 0.3816(1)         & 0.3478(1)                   && 0.3822(1)         & 0.3524(1)                     \\
$r$ (side)                               & 0.3713(1)         & 0.2495(1)                   && 0.3719(1)         & 0.2472(1)                   && 0.3719(1)         & 0.2508(1)                     \\
$r$ (back)                               & 0.3767(1)         & 0.2820(1)                   && 0.3773(1)         & 0.2796(1)                   && 0.3775(1)         & 0.2833(1)                     \\
$r$ (volume)$\rm ^a$                     & 0.3695            & 0.2579                      && 0.3701            & 0.2556                      && 0.3702            & 0.2592                        \\ [1.0mm]
\multicolumn{6}{l}{Third-body parameters:}                                                                                                                                                         \\        
$a^{\prime}$($R_\odot$)                  & \multicolumn{2}{c}{564(6)}                      && \multicolumn{2}{c}{547(4)}                      && \multicolumn{2}{c}{547(4)}                        \\        
$i^{\prime}$ (deg)                       & \multicolumn{2}{c}{76.8}                        && \multicolumn{2}{c}{76.8}                        && \multicolumn{2}{c}{76.8}                          \\        
$e^{\prime}$                             & \multicolumn{2}{c}{0.50(8)}                     && \multicolumn{2}{c}{0.46(6)}                     && \multicolumn{2}{c}{0.47(6)}                       \\        
$\omega^{\prime}$  (deg)                 & \multicolumn{2}{c}{344(11)}                     && \multicolumn{2}{c}{313(10)}                     && \multicolumn{2}{c}{303(10)}                       \\        
$P^{\prime}$ (d)                         & \multicolumn{2}{c}{1017(17)}                    && \multicolumn{2}{c}{970(11)}                     && \multicolumn{2}{c}{970(11)}                       \\        
$T_{\rm c}^{\prime}$ (BJD)               & \multicolumn{2}{c}{2,455,330(30)}               && \multicolumn{2}{c}{2,457,327(37)}               && \multicolumn{2}{c}{2,457,355(40)}                 \\ [1.0mm]
\multicolumn{6}{l}{Spot parameters:}                                                                                                                                                               \\        
Colatitude (deg)                         & \dots             & \dots                       && 85(3)             & \dots                       && \dots             & 75(2)                         \\        
Longitude (deg)                          & \dots             & \dots                       && 249(1)            & \dots                       && \dots             & 288(1)                        \\        
Radius (deg)                             & \dots             & \dots                       && 13(3)             & \dots                       && \dots             & 27(2)                         \\        
$T$$\rm _{spot}$/$T$$\rm _{local}$       & \dots             & \dots                       && 1.03(2)           & \dots                       && \dots             & 0.90(2)                       \\
$\Sigma W(O-C)^2$                        & \multicolumn{2}{c}{0.0025}                      && \multicolumn{2}{c}{0.0020}                      && \multicolumn{2}{c}{0.0020}                        \\ [1.0mm]
\multicolumn{6}{l}{Absolute parameters:}                                                                                                                                                           \\            
$M$($M_\odot$)                           & 1.95              &  0.43                       && 1.95              &  0.42                       && 1.95              &  0.44                         \\
$R$($R_\odot$)                           & 1.99              &  1.39                       && 1.99              &  1.37                       && 2.00              &  1.40                         \\
$\log$ $g$ (cgs)                         & 4.13              &  3.79                       && 4.13              &  3.78                       && 4.13              &  3.79                         \\
$\rho$ (g cm$^3)$                        & 0.35              &  0.23                       && 0.35              &  0.23                       && 0.35              &  0.23                         \\
$L$($L_\odot$)                           & 17.9              &  0.53                       && 17.8              &  0.52                       && 18.0              &  0.54                         \\
$M_{\rm bol}$ (mag)                      & 1.60              &  5.42                       && 1.60              &  5.43                       && 1.59              &  5.41                         \\
BC (mag)                                 & $+$0.01           &  $-$0.89                    && $+$0.01           &  $-$0.88                    && $+$0.01           &  $-$0.88                      \\
$M_{\rm V}$ (mag)                        & 1.59              &  6.31                       && 1.59              &  6.31                       && 1.58              &  6.29                         \\
\enddata
\tablenotetext{a}{Mean volume radius.}
\end{deluxetable}

\begin{deluxetable}{lcrcc}
\tablewidth{0pt}
\tablecaption{Minimum timings determined by the method of Kwee \& van Woerden}
\tablehead{
\colhead{BJD}    & \colhead{Error} & \colhead{Epoch} & \colhead{$O$--$C$} & \colhead{Min}  
}
\startdata
2,454,954.51210  & $\pm$0.00042    & $-$765.0        & $-$0.00180         & I       \\
2,454,954.98364  & $\pm$0.00070    & $-$764.5        & $+$0.00029         & II      \\
2,454,955.45086  & $\pm$0.00034    & $-$764.0        & $-$0.00194         & I       \\
2,454,955.92193  & $\pm$0.00068    & $-$763.5        & $-$0.00032         & II      \\
2,454,956.38994  & $\pm$0.00020    & $-$763.0        & $-$0.00176         & I       \\
2,454,957.32879  & $\pm$0.00031    & $-$762.0        & $-$0.00182         & I       \\
2,454,957.79984  & $\pm$0.00092    & $-$761.5        & $-$0.00022         & II      \\
2,454,958.26783  & $\pm$0.00034    & $-$761.0        & $-$0.00168         & I       \\
2,454,959.67758  & $\pm$0.00073    & $-$759.5        & $-$0.00029         & II      \\
2,454,960.14545  & $\pm$0.00065    & $-$759.0        & $-$0.00187         & I       \\
\enddata
\tablecomments{This table is available in its entirety in machine-readable and Virtual Observatory (VO) forms in the online journal. 
A portion is shown here for guidance regarding its form and content.}
\end{deluxetable}

\begin{deluxetable}{lcccrc}
\tablewidth{0pt}
\tablecaption{Minimum timings determined by the W-D code}
\tablehead{
\colhead{BJD}    & \colhead{Error} & \colhead{Epoch} & \colhead{$O$--$C_{\rm full}$} 
}
\startdata
2,454,953.57307  & $\pm$0.00020    & $-$766          & $-$0.00004     \\
2,454,954.51230  & $\pm$0.00016    & $-$765          & $+$0.00027     \\
2,454,955.45105  & $\pm$0.00019    & $-$764          & $+$0.00011     \\
2,454,956.38984  & $\pm$0.00021    & $-$763          & $-$0.00000     \\
2,454,957.32878  & $\pm$0.00017    & $-$762          & $+$0.00002     \\
2,454,958.26782  & $\pm$0.00021    & $-$761          & $+$0.00014     \\
2,454,959.20668  & $\pm$0.00023    & $-$760          & $+$0.00009     \\
2,454,960.14566  & $\pm$0.00019    & $-$759          & $+$0.00016     \\
2,454,961.08451  & $\pm$0.00025    & $-$758          & $+$0.00010     \\
2,454,962.02359  & $\pm$0.00023    & $-$757          & $+$0.00027     \\
\enddata
\tablecomments{This table is available in its entirety in machine-readable and Virtual Observatory (VO) forms in the online journal. 
A portion is shown here for guidance regarding its form and content.}
\end{deluxetable}

\begin{deluxetable}{lccccc}
\tablewidth{0pt}
\tablecaption{Parameters for the quadratic {\it plus} LTT ephemeris of KIC 5621294}
\tablehead{
\colhead{Parameter}     & \colhead{Value}              & \colhead{Unit}   \\
}                                                                                                                                         
\startdata                                                                                                                                
$T_0$                   & 2,455,672.7763783(27)        & BJD              \\
$P$                     & 0.9389051610(31)             & d                \\
$A$                     & $-3.4608(82)\times 10^{-9}$  & d                \\
$a_{\rm b}\sin i_{3}$   & 0.04716(36)                  & au               \\
$e_{\rm b}$             & 0.434(15)                    &                  \\
$\omega_{\rm b}$        & 132.6(1.9)                   & deg              \\
$n_{\rm b}$             & 0.37461(86)                  & deg d$^{-1}$     \\
$T_{\rm b}$             & 2,455,163.4(4.2)             & BJD              \\
$P_{\rm b}$             & 961.0(2.2)                   & d                \\
$K_{\rm b}$             & 22.49(17)                    & s                \\
$f(M_{3})$              & 0.00001515(12)               & M$_\odot$        \\ \hline 
$M_{3} \sin i_{3}$      & 0.04478(17)                  & M$_\odot$        \\
$a_{3} \sin i_{3}$      & 2.5168(48)                   & au               \\
$e_3$                   & 0.434(15)                    &                  \\
$\omega_{3}$            & 312.6(1.9)                   & deg              \\
$P_3$                   & 961.0(2.2)                   & d                \\[0.5mm]
rms scatter             & 10.88                        & s                \\
$\chi^2 _{\rm red}$     & 1.107                        &                  \\
\enddata
\end{deluxetable}

\end{document}